\documentclass[journal=ancac3,manuscript=article]{achemso}

\usepackage[version=3]{mhchem} % Formula subscripts using \ce{}

\author{Adrian G. Swartz}
\author{Patrick M. Odenthal}
\affiliation{Department of Physics and Astronomy, University of California, Riverside CA 92521}
\author{Yufeng Hao}
\author{Rodney S. Ruoff}
\affiliation{Department of Mechanical Engineering and the Texas Materials Institute, The University of Texas at Austin, Austin TX 78712}
\author{Roland K. Kawakami}
\affiliation{Department of Physics and Astronomy, University of California, Riverside CA 92521}
\email{roland.kawakami@ucr.edu}

\title{Integration of the Ferromagnetic Insulator EuO onto Graphene}


\begin{document}
%%%%%%%%%%%%%%%%%%%%%%%%%%%%%%%%%%%%%%%%%%%%%%%%%%%%%%%%%%%%%%%%%%%%%
%% The manuscript does not need to include \maketitle, which is
%% executed automatically.  The document should begin with an
%% abstract, if appropriate.  If one is given and should not be, the
%% contents will be gobbled.
%%%%%%%%%%%%%%%%%%%%%%%%%%%%%%%%%%%%%%%%%%%%%%%%%%%%%%%%%%%%%%%%%%%%%
\begin{abstract}
	We have demonstrated the deposition of EuO films on graphene by reactive molecular beam epitaxy in a special adsorption-controlled and oxygen-limited regime, which is a critical advance toward the realization of the exchange proximity interaction (EPI). It has been predicted that when the ferromagnetic insulator (FMI) EuO is brought into contact with graphene, an overlap of electronic wavefunctions at the FMI/graphene interface can induce a large spin splitting inside the graphene. Experimental realization of this effect could lead to new routes for spin manipulation, which is a necessary requirement for a functional spin transistor. Furthermore, EPI could lead to novel spintronic behavior such as controllable magnetoresistance, gate tunable exchange bias, and quantized anomalous Hall effect. However, experimentally, EuO has not yet been integrated onto graphene.  Here we report the successful growth of high quality crystalline EuO on highly-oriented pyrolytic graphite (HOPG) and single-layer graphene. The epitaxial EuO layers have (001) orientation and do not induce an observable D peak (defect) in the Raman spectra. Magneto-optic measurements indicate ferromagnetism with Curie temperature of 69 K, which is the value for bulk EuO. Transport measurements on exfoliated graphene before and after EuO deposition indicate only a slight decrease in mobility.
\end{abstract}


\paragraph{}The exchange proximity interaction (EPI) has been predicted to exist at the interface between a ferromagnetic insulator (FMI) and graphene, originating from an overlap of electronic wavefunctions\cite{Haugen:2008,Semenov:2007}. In particular, the ferromagnetic insulator EuO has been estimated theoretically to induce a spin splitting in graphene of the order 5 meV\cite{Haugen:2008}. EPI has been suggested for novel spintronic device functionality in a wide variety of applications such as induced magnetism in graphene\cite{Haugen:2008,Yokoyama:2008,Soodchomshom:2008,Dell'Anna:2009}, controllable magnetoresistance\cite{Haugen:2008,Semenov:2008a,Zou:2009,Yu:2011}, gate tunable manipulation of spin transport\cite{Semenov:2007,Michetti:2010}, gate tunable exchange bias\cite{Semenov:2008b}, spin transfer torque\cite{Zhou:2010,Yokoyama:2011}, as well as being a necessary requirement for the observation of the quantized anomalous Hall effect in graphene\cite{Qiao:2010,Tse:2011}. While theoretical predictions have been numerous, EPI at the FMI/graphene interface has yet to be experimentally observed.

EuO is a model FMI as it exemplifies an isotropic Heisenberg ferromagnet\cite{Mauger:1986}. EuO has a half filled 4{\it{f}} shell which determines the magnetic properties leading to a magnetization of 7 Bohr magnetons per Eu atom. Further, because the 4{\it{f}} shell is electrically inert due to its localized orbitals, the unoccupied exchange split 5{\it{d}} band governs the charge transport characteristics and exchange overlap in stoichiometric EuO. However, part of the reason that EPI has yet to be observed in the EuO/graphene system is due to the difficulty in materials synthesis of high quality stoichiometric EuO thin films. EuO is not thermodynamically stable and readily converts to nonmagnetic Eu$_2$O$_3$\cite{Samsanov:1982}. Furthermore, oxygen deficient EuO$_{1-x}$ exhibits a metal to insulator transition\cite{Oliver:1972} with a conductive ferromagnetic phase\cite{Schmehl:2007,Steeneken:2002}. In typical materials synthesis techniques such as reactive molecular beam epitaxy (MBE), maintaining stoichiometry by flux matching generally leads to the formation of either Eu$_2$O$_3$ or EuO$_{1-x}$. In order to possibly realize EPI in graphene, a critical first step is the integration of high quality stoichiometric EuO thin films with graphene.

Only recently, through the development of a special growth regime, have reliable stoichiometric films been readily produced\cite{Steeneken:2002,Ulbricht:2008,Sutarto:2009,Swartz:2010,Swartz:2012}. The regime can be understood as follows: a high-purity elemental Eu flux is incident upon a heated substrate maintained at a temperature for which the incident Eu atoms re-evaporate off the substrate surface (i.e. distillation). Notably, distillation is highly substrate dependent and works well on certain oxides\cite{Sutarto:2009,Swartz:2012}, but fails in the case of direct growth on GaAs\cite{Swartz:2010}. Once distillation is achieved, the introduction of a small oxygen partial pressure allows for the formation of EuO while excess Eu atoms are re-evaporated. This ensures proper stoichiometry of the EuO film\cite{Steeneken:2002,Ulbricht:2008,Sutarto:2009,Swartz:2010,Swartz:2012}. If the oxygen partial pressure is increased, the EuO growth rate increases until a critical O$_2$ pressure is reached and Eu$_2$O$_3$ forms. In this way, the growth rate is determined by the oxygen pressure and is termed adsorption-controlled (distillation) and oxygen-limited. To date there is no evidence that stoichiometric EuO can be integrated with {\it{sp$^2$}} bonded carbon based materials.

In this study, we employ reactive MBE to investigate the deposition of EuO thin films onto graphene. First, we examine the viability of Eu distillation for {\it{sp$^2$}} bonded carbon materials by examining highly-oriented pyrolitic graphite (HOPG) substrate, which allows for standard thin film characterization techniques such as Auger spectroscopy, reflection high energy electron diffraction (RHEED), and x-ray diffraction (XRD). Within the distillation and oxygen-limited regime, stoichiometric EuO(001) is shown to grow epitaxially on HOPG(0001) substrate. Such films are shown to be uniform and flat by {\it{ex-situ}} atomic force microscopy (AFM). Further, EuO is integrated onto mechanically exfoliated graphene flakes as well as large area graphene grown by chemical vapor deposition (CVD). Raman spectroscopy after EuO deposition on exfoliated graphene exhibits the absence of a D peak, indicating that, despite the high temperatures of deposition, EuO thin films do not induce significant defects to the underlying graphene. This is supported by four point resistivity measurements that indicate only a slight reduction of mobility. Also, we investigate the magnetic properties of EuO on HOPG and CVD graphene and find a Curie temperature (T$_\text{C}$) of 69 K, the bulk EuO value. This advance in materials synthesis allows for future studies of EPI at FMI/graphene interfaces.

\subsection{Results and discussion}
\begin{figure}
\includegraphics{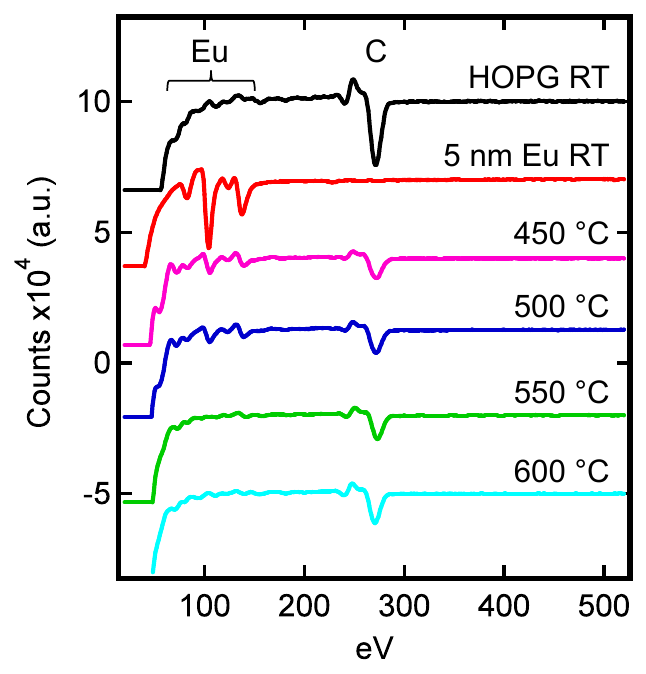}
\caption{\label{figure:Fig. 1} Auger spectroscopy of Eu deposited on HOPG at several different substrate temperatures. Pristine HOPG (black) shows a carbon peak at 272 eV. 5nm Eu deposited at room temperature (red) shows Eu peaks only. At higher growth temperatures a combination of Eu and C peaks are present. Above 550 $\,^{\circ}\text{C}$ there is no evidence of Eu in the spectrum.}
\end{figure}

\paragraph{}First, we establish the growth parameters by investigating EuO growth on HOPG using Auger spectroscopy, RHEED, and XRD. Fresh surfaces of HOPG (SPI, grade ZYA) are obtained by peeling with 3M scotch tape and subsequently loaded into the UHV growth chamber and annealed at 600 $\,^{\circ}\text{C}$ for 30 min. Auger spectroscopy for a pristine HOPG surface is shown in Fig.~\ref{figure:Fig. 1} (top curve). The spectrum is characterized by a peak at 272 eV identifying carbon. Since the temperature required for efficient Eu distillation is highly substrate dependent, we cannot rely on previous results for distillation temperatures based on oxide substrates\cite{Steeneken:2002,Ulbricht:2008,Sutarto:2009,Swartz:2012}. Therefore, we first investigated the optimal re-evaporation temperature on HOPG.  Without introducing a partial pressure of molecular oxygen, an incident Eu flux (8-9 \AA/min.) is introduced to the substrate, which is maintained at a fixed temperature. Fig.~\ref{figure:Fig. 1} shows Auger spectra for Eu metal deposited at room temperature (RT), 450 $\,^{\circ}\text{C}$, 500 $\,^{\circ}\text{C}$, 550 $\,^{\circ}\text{C}$, and 600 $\,^{\circ}\text{C}$. For each substrate temperature, Eu is deposited for the time equivalent to produce a 5 nm Eu film at RT. Eu Auger peaks at 83, 104, 124, and 138 eV can be seen in the RT spectrum of Fig.~\ref{figure:Fig. 1}. As the substrate temperature is increased, the relative peak height of Eu to C decreases indicating a smaller amount of Eu material on HOPG. This indicates the onset of re-evaporation of the Eu atoms. For the case of 550 $\,^{\circ}\text{C}$ and 600 $\,^{\circ}\text{C}$, the Auger spectra shows only the carbon peak at 272 eV and no evidence of Eu material. Therefore, full distillation of Eu on HOPG is achieved above 550 $\,^{\circ}\text{C}$.

\begin{figure}
\includegraphics{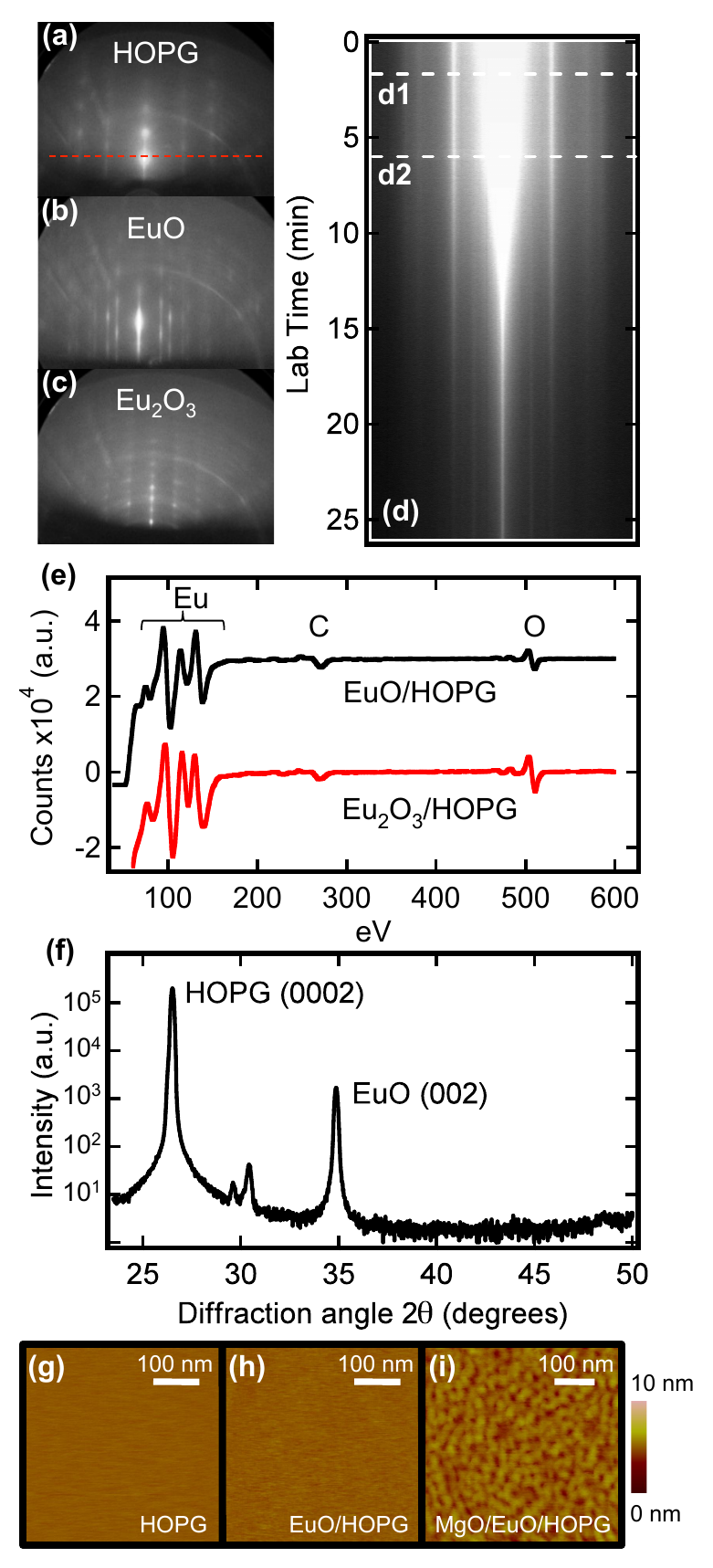}
\caption{\label{figure:Fig. 2} Characterization of EuO thin films on HOPG. \textbf{(a)-(c)} RHEED patterns for pristine HOPG, 5nm EuO deposited on HOPG, and Eu$_2$O$_3$ on HOPG. \textbf{(d)} Time evolution of dashed red curve in \textbf{(a)}. Dashed line d1 indicates the opening of the Eu shutter and incidence of Eu flux to the substrate. Dashed line d2 signifies the introduction of molecular oxygen into the chamber. \textbf{(e)} Auger spectroscopy of EuO/HOPG and Eu$_2$O$_3$/HOPG. \textbf{(f)} XRD $\theta$-2$\theta$ scan showing the HOPG (0002) peak and EuO (002) peak. \textbf{(g)}-\textbf{(i)} {\it{ex-situ}} AFM scans for peeled HOPG, EuO(5 nm)/HOPG(0001), and MgO(2 nm)/EuO(5 nm)/HOPG(0001) in order from left to right.}
\end{figure}

Once in the distillation regime, the introduction of an oxygen flux smaller than the elemental Eu flux should produce stoichiometric EuO films. We investigate the formation of EuO on HOPG substrate by maintaining the substrate at 550 $\,^{\circ}\text{C}$ for distillation and then introduce a molecular oxygen partial pressure (P$_{\text{O}_2}=1.0\times10^{-8}$ Torr) into the UHV system. {\it{In-situ}} RHEED images probe the sample surface crystalline structure. Fig.~\ref{figure:Fig. 2} (a) and (b) show the RHEED patterns for HOPG and the EuO layer after 5 nm of growth, respectively. The RHEED pattern for the HOPG substrate is unaltered upon in-plane rotation. This is expected since HOPG has out-of-plane (0001) orientation but has in-plane rotational disorder. The RHEED pattern of the EuO layer shows double streak features and in-plane rotation has no effect on the RHEED pattern, similar to the HOPG substrate. Examination of the EuO RHEED diffraction rods indicates EuO(001) with a superposition of both [100] and [110] in-plane orientations\cite{Swartz:2012}. We can better understand the growth evolution of the EuO film by examining the time lapse of a line cut of the RHEED pattern. A typical line cut, as depicted in Fig.~\ref{figure:Fig. 2} (a) (red dashed line), samples the intensity of several diffraction rods across the RHEED pattern. Fig.~\ref{figure:Fig. 2} (d) displays the time evolution of a line cut for EuO growth on HOPG in the distillation and oxygen-limited regime. Between 0 min.~and dashed line d1, the high intensity streaks correspond to the diffraction rods as seen in Fig.~\ref{figure:Fig. 2} (a) of the pristine HOPG pattern. Dashed line d1 indicates the introduction of Eu flux, during which time the HOPG diffraction rods remain unchanged as Eu re-evaporates off the HOPG surface. A partial pressure of oxygen (P$_{\text{O}_2}=1.0\times10^{-8}$ Torr) is leaked into the chamber at dashed line d2. The subsequent time evolution shows a smooth transition from HOPG streaks to EuO indicating epitaxial growth.

Eu$_2$O$_3$ can be grown by increasing the O$_2$ partial pressure to $3\times10^{-7}$ Torr. The Eu$_2$O$_3$ RHEED pattern is displayed in Fig.~\ref{figure:Fig. 2} (c) and shows a clear distinction from the oxygen-limited growth which produces EuO.  We further investigate the difference between the two oxygen regimes by looking at their respective Auger data as shown in Fig.~\ref{figure:Fig. 2} (e). An oxygen peak at 510 eV is present for both samples. To our knowledge this is the first report on Auger spectroscopy for EuO. Taking the Eu 104 eV peak, EuO has a Eu:O peak ratio of 6.63, while Eu$_2$O$_3$ has a Eu:O ratio of 4.34 indicating increased oxygen content in Eu$_2$O$_3$. Comparison of the ratios 6.63/4.34 = 1.53 to the expected Eu:O/Eu$_2$:O$_3$ = 1.5 is in close agreement. However, it should be noted that while this analysis is useful in verifying oxygen content between the two growths (i.e. EuO vs. Eu$_2$O$_3$), it is not sufficient for determining precise stoichiometry of the EuO oxidation state.

{\it{Ex-situ}} XRD $\theta$-2$\theta$ scans, Fig.~\ref{figure:Fig. 2} (f), serve to elucidate the structure of EuO deposited on HOPG. For XRD measurements, approximately 50 nm EuO was grown on HOPG and capped with 3 nm polycrystalline Al. A clear EuO (002) peak is seen in the $\theta$-2$\theta$ scan and there are no other peaks associated with another EuO orientation indicating that entire EuO film is oriented (001), in agreement with the RHEED analysis. There are no detectable peaks associated with Eu$_2$O$_3$. There are two small peaks at 29.63 $^\circ$ and 30.43 $^\circ$ associated with Eu$_3$O$_4$ (040) and (320), possibly due to oxidation through the thin capping layer.

It is generally expected that FCC materials (EuO, Ni, etc..) would favor (111) orientations with hexagonal materials due to the surface symmetry. However, the RHEED and XRD data clearly indicate the orientation EuO(001)/HOPG(0001) is preferred. In the absence of other factors, the orientation preference may be partly explained by the lattice mismatch between EuO and graphene. EuO has a bulk lattice constant of 0.514 nm and 0.246 nm for graphite, leading to a lattice mismatch of 4.3\% for EuO(001)/HOPG(0001) growth orientations. The mismatch for EuO(111)/HOPG(0001) is either 10\% or 17\% depending on the ratio of relative lattice spacings (i.e. 1:4 or 1:3 for EuO:graphene). However, while mismatch considerations might suggest a favorable orientation, it cannot explain the lack of symmetry between the rock salt surface and graphene. Previous work\cite{Swartz:2012} has shown lattice mismatch to be less of a key factor for EuO epitaxy than other growth concerns. Surface energies, which are lowest for (100) rock salt surfaces\cite{Henrich:1996}, are likely relevant and may provide a possible explanation for the observed growth orientation.

To investigate the surface morphology of the EuO films, we have performed {\it{ex-situ}} AFM on peeled HOPG(0001) substrate, a EuO(5 nm)/HOPG(0001) film, and a MgO(2 nm)/EuO(5 nm)/HOPG(0001) bilayer. The resulting AFM scans are displayed in Fig. 2 (g), (h), and (i), with rms roughness values 0.1 nm, 0.2 nm, and 0.5 nm, respectively. It must be noted that the EuO surface is likely oxidized to Eu$_2$O$_3$ during the {\it{ex-situ}} measurement. In any case, the scans clearly show that the films are uniform, relatively flat, and pinhole free. This is crucial for possible use as a gate dielectric.

\begin{figure}
\includegraphics{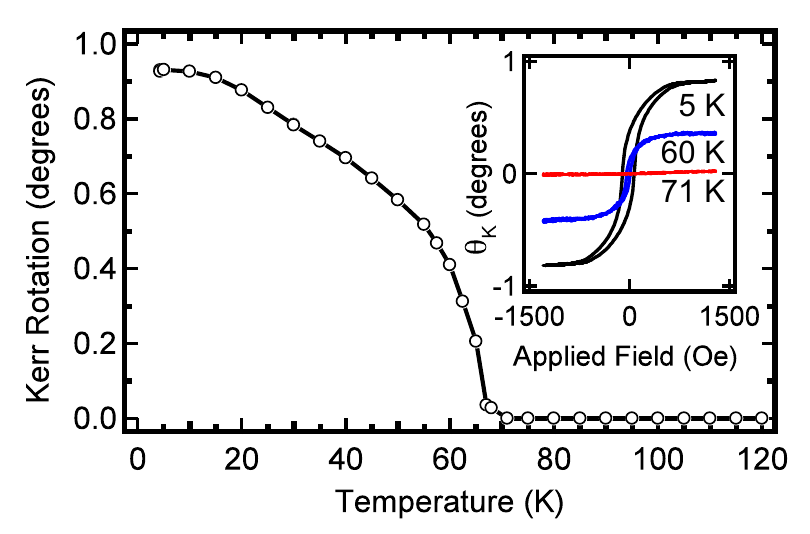}
\caption{\label{figure:Fig. 3} Temperature dependence of the magnetization of EuO/HOPG measured by the Magneto-optic Kerr effect (MOKE). Inset shows several characteristic hysteresis loops at different temperatures. }
\end{figure}

Due to EuO's large magneto-optic response\cite{Ahn:1970}, the magneto-optic Kerr effect (MOKE) serves as a sensitive probe of the magnetic behavior of the sample. Linearly polarized light is reflected off the sample surface and the resulting polarization rotates an amount, $\theta_\text{K}$, which is proportional to the magnetization of the film. The sample structure is Al(2nm)/EuO(5nm)/HOPG(0001) and is measured in an optical flow cryostat separate from the UHV growth chamber. Fig.~\ref{figure:Fig. 3} inset shows magnetic hysteresis loops measured at 5 K, 60 K, and 71 K. At 5 K, the remanence (M$_\text{R}$/M$_\text{S}$) is 0.37, the coercive field (H$_\text{C}$) is 87 Oe, and the saturation Kerr rotation is 0.93 degrees. Fig.~\ref{figure:Fig. 3} shows a temperature dependence of the saturation magnetization with T$_\text{C}$ = 69 K, the bulk value for EuO.

\begin{figure}
\includegraphics{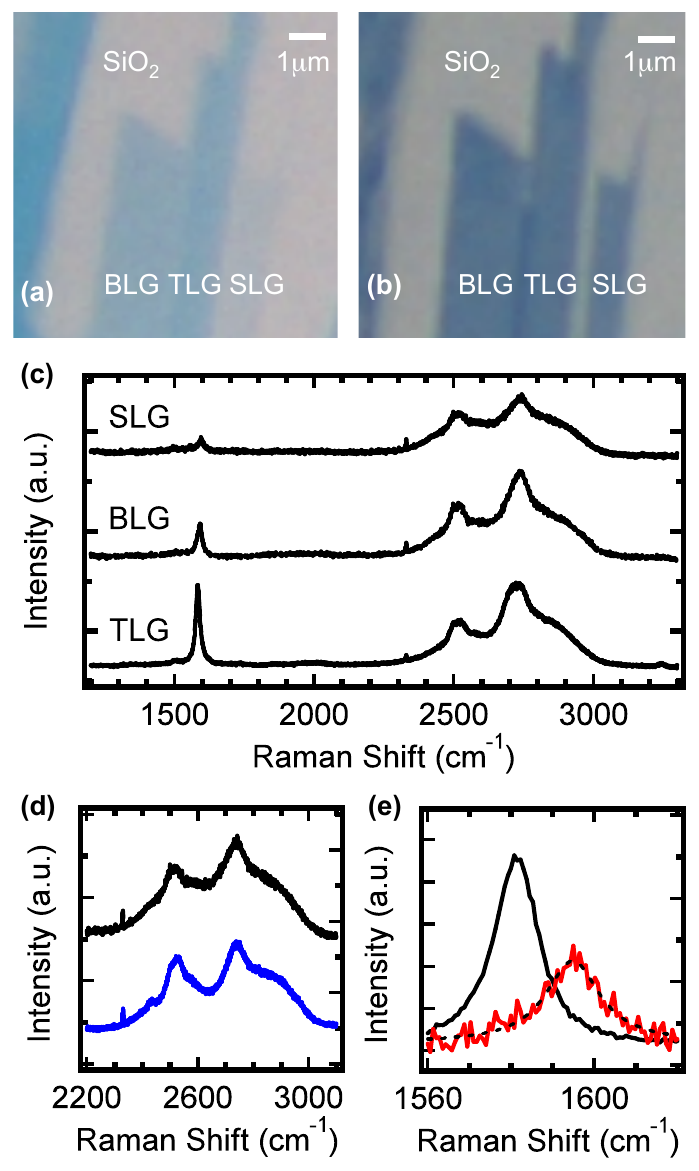}
\caption{\label{figure:Fig. 4} EuO thin films deposited on exfoliated graphene flakes. \textbf{(a)} Optical microscope image of SLG, BLG, and TLG on SiO$_2$/Si substrate before EuO deposition. The scale bar indicates 1 $\mu$m. \textbf{(b)} Optical microscope image of of the same sample after EuO deposition. \textbf{(c)} Raman spectroscopy of EuO deposited on single layer, bilayer, and trilayer graphene. \textbf{(d)} Raman spectroscopy of EuO/SLG (black) compared to EuO/YSZ(001) (blue) in the region around the 2D peak. \textbf{(e)} Raman spectroscopy of the G band for pristine SLG (black) compared to EuO/SLG (red) with the intensities scaled for ease of viewing. Dashed black line is a lorentzian fit to the G peak for EuO deposited on SLG.}
\end{figure}

While EuO/HOPG serves as a useful system for examining the epitaxy of EuO on {\it{sp$^2$}} bonded carbon and allows for the use of standard thin film characterization techniques, realization of EPI at EuO/graphene interfaces requires direct integration of EuO on either exfoliated or CVD graphene. Graphene flakes are mechanically exfoliated onto 300 nm SiO$_2$/Si substrate using standard techniques\cite{Novoselov:2004}. Single layer (SLG), bilayer (BLG), and trilayer (TLG) flakes are identified under an optical microscope and confirmed by Raman spectroscopy\cite{Hao:2010}. A 5 nm EuO film is deposited on top of exfoliated graphene flakes on SiO$_2$ and capped with 2 nm MgO. Fig.~\ref{figure:Fig. 4} (a) shows an optical microscope image of pristine graphene flakes while Fig.~\ref{figure:Fig. 4} (b) shows the same flakes after EuO deposition with noticeable darkening of the graphene flakes. Raman spectroscopy (535 nm laser) of EuO/graphene for several flake thicknesses is shown in Fig.~\ref{figure:Fig. 4} (c). Several key features are immediately apparent for EuO deposited onto graphene flakes. First, we do not observe a D peak above the noise level of the measurement. The D peak is typically associated with induced disorder\cite{Ferrari:2006,Gupta:2006,Malard:2009} suggesting that the deposition process does not induce significant defects when compared with reports for oxide growth by PLD, e-beam, and sputter deposition\cite{Tang:2010}. Second, the G peak shrinks in relative size compared with features above 2200 cm$^{-1}$ due to decreased signal from the impeding EuO overlayer. Lastly, the spectra exhibit a significant modification around the graphene 2D peak. To better understand this behavior, we compare with single crystal EuO on lattice-matched YSZ(001)\cite{Sutarto:2009,Swartz:2010}. Fig.~\ref{figure:Fig. 4} (d) shows the Raman spectra for EuO/SLG and EuO/YSZ around the graphene 2D peak. The features are nearly identical, indicating they are not related to graphene phonon modes.

Raman spectroscopy is a useful technique for investigating external effects on graphene such as doping, strain, and defects\cite{Gupta:2006,Malard:2009,Tang:2010,Pisana:2007,Das:2008,Ni:2008a,Ni:2008b,Ding:2010,Cheng:2011}. A closer examination of the graphene G peak for 5 nm EuO deposited on SLG (Fig.~\ref{figure:Fig. 4} (e)) shows blue shifting of the G peak by 14 cm$^{-1}$, from 1581 cm$^{-1}$ to 1595 cm$^{-1}$. We also note blue shifts for EuO/BLG and EuO/TLG of 10 cm$^{-1}$ and 6 cm$^{-1}$, respectively. Both charge doping and induced strain could possibly explain the blue shifted G peak after EuO deposition\cite{Pisana:2007,Das:2008,Ni:2008a,Ni:2008b,Ding:2010,Cheng:2011}. For SLG, while a shift of 14 cm$^{-1}$ would suggest an induced charge doping greater than $6\times10^{12}$ cm$^{-2}$, an increase in the carrier concentration of that magnitude is expected to decrease the FWHM by approximately 8 cm$^{-1}$\cite{Pisana:2007}. Interestingly, for EuO deposited on graphene flakes, the FWHM of the G peak is 13 cm$^{-1}$ for pristine graphene and 15 cm$^{-1}$ with EuO, making charge doping unlikely as the sole cause of the G peak shift. Alternatively, the shift could be caused by strain and is comparable to that reported for annealed SiO$_2$/graphene/SiO$_2$\cite{Ni:2008a}. The 2D peak would shed light on this issue, but is not accessible due to the EuO overlayer.

\begin{figure}
\includegraphics{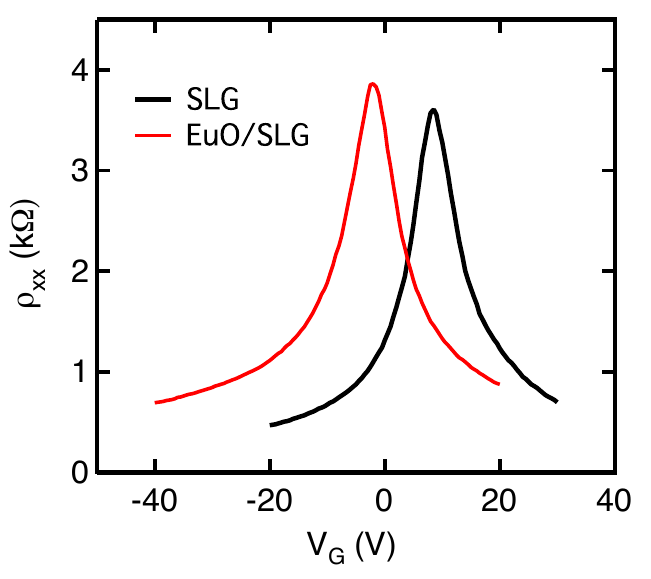}
\caption{\label{figure:Fig. 5}  Gate dependent resistivity for pristine graphene and for the same device with MgO(2 nm)/EuO(2 nm) overlayer.}
\end{figure}

Next, we discuss the effect of an EuO overlayer on charge transport.  Graphene devices are fabricated using standard e-beam lithography techniques with Ti/Au (10 nm/60 nm) electrodes\cite{Pi:2009}.  The resistivity is measured using 1 $\mu$A excitation at 11 Hz AC for lock-in detection in a four point geometry. Fig.~\ref{figure:Fig. 5} shows the resistivity for pristine SLG (black curve) with charge neutrality point at V$_{CNP}$ = 8 V.  The device is then loaded into the MBE chamber for growth of 2 nm EuO followed by a 2 nm MgO capping layer. The charge neutrality point after growth (red curve) is V$_{CNP}$ = -2 V. The electron mobility can be determined from the slope of the conductivity ($\mu=\Delta\sigma/e\Delta n$).  The carrier concentration, $n$, is determined from the relation $n=-\alpha(V_G-V_{CNP})$, where 
$\alpha=7.2\times10^{10}$ V$^{-1}$cm$^{-2}$ for 300 nm SiO$_2$ gate dielectric. The resulting electron mobility for pristine SLG and EuO/SLG are $\mu_e=4600$ cm$^2$/Vs and $\mu_e=4080$ cm$^2$/Vs, respectively. Thus, the deposition of EuO on the graphene surface, does not significantly decrease the mobility.

\begin{figure}
\includegraphics{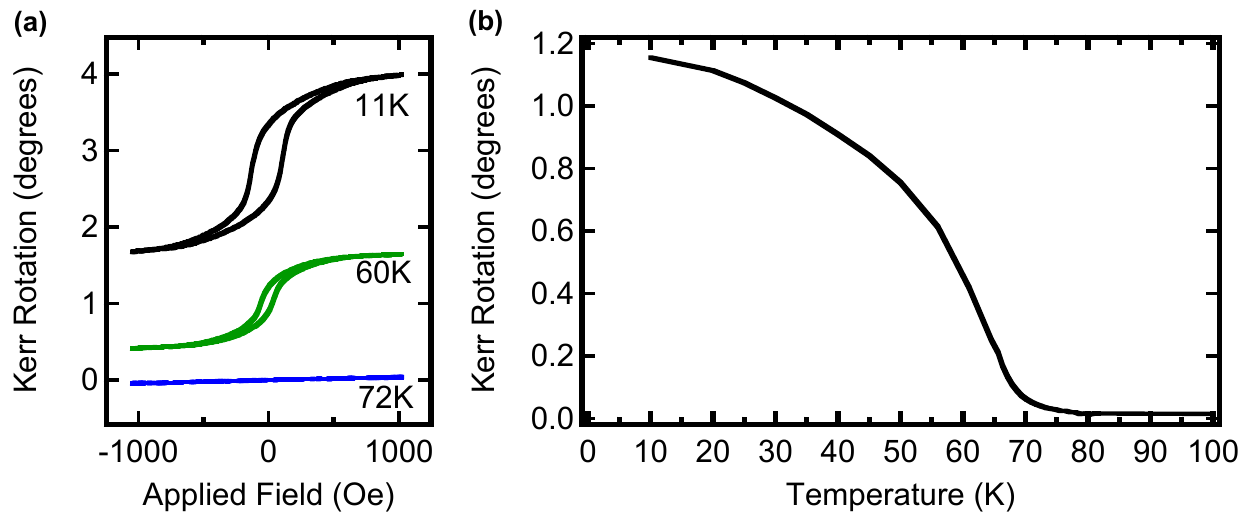}
\caption{\label{figure:Fig. 6} \textbf{(a)} Several characteristic hysteresis loops at different temperatures for 5 nm EuO deposited  onto CVD graphene on SiO$_2$/Si. \textbf{(b)} Temperature dependence of the magnetization measured in degrees.}
\end{figure}

Lastly, we investigate the magnetic properties of EuO/graphene. For this, we employ large-area graphene which has been demonstrated to produce high-quality films with large grains\cite{Li:2009}, and is therefore desirable for MOKE characterization which has a spot size with $\sim$40 $\mu$m diameter. Large area graphene is grown by chemical vapor deposition on copper foil and subsequently transferred to SiO$_2$/Si\cite{Li:2009}. Next, 5 nm EuO thin film with 2 nm MgO capping layer is deposited on the CVD graphene in the distillation and oxygen-limited regime. Fig.~\ref{figure:Fig. 6} (a) shows several MOKE hysteresis loops taken at 11 K, 60 K, and 72 K. As typical with EuO thin films, we observe a large Kerr rotation above 1 degree, which subsequently decreases in magnitude as the temperature is increased towards the Curie temperature of 69 K as shown in Fig.~\ref{figure:Fig. 6} (b).

\subsection{Conclusion}
\paragraph{}We have investigated the integration of the ferromagnetic insulator EuO with graphene. Using Auger spectroscopy, we find that distillation (re-evaporation) of Eu from the graphene surface occurs for temperatures above 550 $\,^{\circ}\text{C}$. Employing the distillation and oxygen-limited regime, EuO was deposited on HOPG and graphene. The structural, chemical, and magnetic properties of these heterostructures were investigated by RHEED, XRD, AFM, Raman, Auger, and MOKE. EuO films grow epitaxially on honeycomb carbon with (001) orientation and the EuO does not induce significant defects in the exfoliated graphene. The growth technique presented here, demonstrates a significant materials advance in the field of oxide growth on graphene, which is notoriously difficult due to the chemically inert nature of the {\it{sp$^2$}} surface. EuO films exhibit ferromagnetism with a Curie temperature of 69 K, equal to the bulk value. The excellent structural and magnetic properties combined with the direct integration without the aid of a buffer layer is a key advance towards experimental observation of the exchange proximity effect at the EuO/graphene interface. 

\subsection{Experimental Methods}

Elemental europium metal (99.99\%) is evaporated from a low temperature thermal cell. After proper degassing, a Eu background pressure below $4\times10^{-9}$ Torr is maintained for rates between 8-9 \AA/min. Molecular oxygen (99.999\%) is leaked into the chamber and the partial pressure is determined by leaking in an amount P$_{\text{O}_2}$ above the background pressure as measured by an ion gauge. Typically, a partial oxygen pressure of $1\times10^{-8}$ Torr is used for which 30 min.~growth time produces films approximately 5 nm thick\cite{Swartz:2010}. The substrate temperature is monitored by a thermocouple located on the platen face. The UHV MBE chamber has a base pressure of \mbox{$\sim1\times10^{-10}$ Torr} and is equipped with {\it{in-situ}} RHEED. Samples are transferred to an adjacent chamber for 3 keV Auger spectroscopy with a base pressure less than $5\times10^{-9}$ Torr. XRD measurements were performed at UCSB MRL Central Facilities. Longitudinal MOKE is performed in an optical flow cryostat with a {\it{p}}-polarized laser beam (635 nm) and an incidence angle of 45 degrees with respect to the in-plane magnetization direction. The laser intensity is 100 $\mu$W focused to a spot size of $\sim$40 $\mu$m in diameter.  Large-area graphene is produced by low pressure CVD as reported by Li, \emph{et al.}\cite{Li:2009}. 25 $\mu$m thick Cu foil (Alfa Aesar, item No. 13382) is loaded into a tube furnace and heated to 1035 $\,^{\circ}\text{C}$. After a 10 min.~anneal in H$_2$ with a flow rate of 2 sccm and pressure, P$_\text{furnace}$= \mbox{$2.5\times10^{-2}$} mbar, 7 sccm of CH$_4$ is introduced for a total pressure of \mbox{$1.4\times10^{-1}$} mbar. After cooling down and removal from the furnace, the Cu is etched away with iron nitrate and transferred onto SiO$_2$/Si substrate with the aid of poly-methyl methacrylate (PMMA) as mechanical support. The PMMA is removed with acetone at room temperature and followed by IPA cleaning. Before EuO growth, the large area graphene sample is annealed at 600 $\,^{\circ}\text{C}$ under UHV condition.

\begin{acknowledgement}

We thank Jared Wong and Renjing Zheng for technical assistance. This work was funded by ONR (N00014-12-1-0469). YH and RSR acknowledge support from NRI-SWAN and ONR (N00014-10-1-0254).

\end{acknowledgement}

%\begin{suppinfo}

%This will usually read something like: ``Experimental procedures and
%characterization data for all new compounds. The class will
%automatically add a sentence pointing to the information on-line:

%\end{suppinfo}

\providecommand*\mcitethebibliography{\thebibliography}
\csname @ifundefined\endcsname{endmcitethebibliography}
  {\let\endmcitethebibliography\endthebibliography}{}

%% graphical table of contents.
%%%%%%%%%%%%%%%%%%%%%%%%%%%%%%%%%%%%%%%%%%%%%%%%%%%%%%%%%%%%%%%%%%%%%


\begin{mcitethebibliography}{37}
\providecommand*\natexlab[1]{#1}
\providecommand*\mciteSetBstSublistMode[1]{}
\providecommand*\mciteSetBstMaxWidthForm[2]{}
\providecommand*\mciteBstWouldAddEndPuncttrue
  {\def\EndOfBibitem{\unskip.}}
\providecommand*\mciteBstWouldAddEndPunctfalse
  {\let\EndOfBibitem\relax}
\providecommand*\mciteSetBstMidEndSepPunct[3]{}
\providecommand*\mciteSetBstSublistLabelBeginEnd[3]{}
\providecommand*\EndOfBibitem{}
\mciteSetBstSublistMode{f}
\mciteSetBstMaxWidthForm{subitem}{(\alph{mcitesubitemcount})}
\mciteSetBstSublistLabelBeginEnd
  {\mcitemaxwidthsubitemform\space}
  {\relax}
  {\relax}

\bibitem[Haugen et~al.(2008)Haugen, Huertas-Hernando, and Brataas]{Haugen:2008}
Haugen,~H.; Huertas-Hernando,~D.; Brataas,~A. Spin Transport in
  Proximity-Induced Ferromagnetic Graphene. \emph{Phys. Rev. B} \textbf{2008},
  \emph{77}, 115406\relax
\mciteBstWouldAddEndPuncttrue
\mciteSetBstMidEndSepPunct{\mcitedefaultmidpunct}
{\mcitedefaultendpunct}{\mcitedefaultseppunct}\relax
\EndOfBibitem
\bibitem[Semenov et~al.(2007)Semenov, Kim, and Zavada]{Semenov:2007}
Semenov,~Y.~G.; Kim,~K.~W.; Zavada,~J.~M. Spin Field Effect Transistor with a
  Graphene Channel. \emph{Appl. Phys. Lett.} \textbf{2007}, \emph{91},
  153105\relax
\mciteBstWouldAddEndPuncttrue
\mciteSetBstMidEndSepPunct{\mcitedefaultmidpunct}
{\mcitedefaultendpunct}{\mcitedefaultseppunct}\relax
\EndOfBibitem
\bibitem[Yokoyama(2008)]{Yokoyama:2008}
Yokoyama,~T. Controllable Spin Transport in Ferromagnetic Graphene Gunctions.
  \emph{Phys. Rev. B} \textbf{2008}, \emph{77}, 073413\relax
\mciteBstWouldAddEndPuncttrue
\mciteSetBstMidEndSepPunct{\mcitedefaultmidpunct}
{\mcitedefaultendpunct}{\mcitedefaultseppunct}\relax
\EndOfBibitem
\bibitem[Soodchomshom et~al.(2008)Soodchomshom, Tang, and
  Hoonsawat]{Soodchomshom:2008}
Soodchomshom,~B.; Tang,~I.-M.; Hoonsawat,~R. Quantum Modulation Effect in a
  Graphene-Based Magnetic Tunnel Junction. \emph{Phys. Lett. A}
  \textbf{2008}, \emph{372}, 5054--5058\relax
\mciteBstWouldAddEndPuncttrue
\mciteSetBstMidEndSepPunct{\mcitedefaultmidpunct}
{\mcitedefaultendpunct}{\mcitedefaultseppunct}\relax
\EndOfBibitem
\bibitem[Dell'Anna and De~Martino(2009)Dell'Anna, and
  De~Martino]{Dell'Anna:2009}
Dell'Anna,~L.; De~Martino,~A. Wave-Vector-Dependent Spin Filtering and Spin
  Transport Through Magnetic Barriers in Graphene. \emph{Phys. Rev. B}
  \textbf{2009}, \emph{80}, 155416\relax
\mciteBstWouldAddEndPuncttrue
\mciteSetBstMidEndSepPunct{\mcitedefaultmidpunct}
{\mcitedefaultendpunct}{\mcitedefaultseppunct}\relax
\EndOfBibitem
\bibitem[Semenov et~al.(2008)Semenov, Zavada, and Kim]{Semenov:2008a}
Semenov,~Y.~G.; Zavada,~J.~M.; Kim,~K.~W. Magnetoresistance in Bilayer Graphene
  \emph{via} Ferromagnet Proximity Effects. \emph{Phys. Rev. B} \textbf{2008},
  \emph{77}, 235415\relax
\mciteBstWouldAddEndPuncttrue
\mciteSetBstMidEndSepPunct{\mcitedefaultmidpunct}
{\mcitedefaultendpunct}{\mcitedefaultseppunct}\relax
\EndOfBibitem
\bibitem[Zou et~al.(2009)Zou, Jin, and qiang Ma]{Zou:2009}
Zou,~J.; Jin,~G.; qiang Ma,~Y. Negative Tunnel Magnetoresistance and Spin
  Transport in Ferromagnetic Graphene Junctions. \emph{J. of Phys.:
  Condens. Matter} \textbf{2009}, \emph{21}, 126001\relax
\mciteBstWouldAddEndPuncttrue
\mciteSetBstMidEndSepPunct{\mcitedefaultmidpunct}
{\mcitedefaultendpunct}{\mcitedefaultseppunct}\relax
\EndOfBibitem
\bibitem[Yu et~al.(2011)Yu, Liang, and Dong]{Yu:2011}
Yu,~Y.; Liang,~Q.; Dong,~J. Controllable Spin Filter Composed of Ferromagnetic
  AB-Stacking Bilayer Graphenes. \emph{Phys. Lett. A} \textbf{2011},
  \emph{375}, 2858 -- 2862\relax
\mciteBstWouldAddEndPuncttrue
\mciteSetBstMidEndSepPunct{\mcitedefaultmidpunct}
{\mcitedefaultendpunct}{\mcitedefaultseppunct}\relax
\EndOfBibitem
\bibitem[Michetti et~al.(2010)Michetti, Recher, and Iannaccone]{Michetti:2010}
Michetti,~P.; Recher,~P.; Iannaccone,~G. Electric Field Control of Spin
  Rotation in Bilayer Graphene. \emph{Nano Lett.} \textbf{2010}, \emph{10},
  4463--4469\relax
\mciteBstWouldAddEndPuncttrue
\mciteSetBstMidEndSepPunct{\mcitedefaultmidpunct}
{\mcitedefaultendpunct}{\mcitedefaultseppunct}\relax
\EndOfBibitem
\bibitem[Semenov et~al.(2008)Semenov, Zavada, and Kim]{Semenov:2008b}
Semenov,~Y.~G.; Zavada,~J.~M.; Kim,~K.~W. Electrical Control of Exchange Bias
  Mediated by Graphene. \emph{Phys. Rev. Lett.} \textbf{2008}, \emph{101},
  147206\relax
\mciteBstWouldAddEndPuncttrue
\mciteSetBstMidEndSepPunct{\mcitedefaultmidpunct}
{\mcitedefaultendpunct}{\mcitedefaultseppunct}\relax
\EndOfBibitem
\bibitem[Zhou et~al.(2010)Zhou, Chen, Wang, Ding, and Zhou]{Zhou:2010}
Zhou,~B.; Chen,~X.; Wang,~H.; Ding,~K.-H.; Zhou,~G. Magnetotransport and
  Current-Induced Spin Transfer Torque in a Ferromagnetically Contacted
  Graphene. \emph{J. of Phys.: Condens. Matter} \textbf{2010},
  \emph{22}, 445302\relax
\mciteBstWouldAddEndPuncttrue
\mciteSetBstMidEndSepPunct{\mcitedefaultmidpunct}
{\mcitedefaultendpunct}{\mcitedefaultseppunct}\relax
\EndOfBibitem
\bibitem[Yokoyama and Linder(2011)Yokoyama, and Linder]{Yokoyama:2011}
Yokoyama,~T.; Linder,~J. Anomalous Magnetic Transport in Ferromagnetic Graphene
  Junctions. \emph{Phys. Rev. B} \textbf{2011}, \emph{83}, 081418\relax
\mciteBstWouldAddEndPuncttrue
\mciteSetBstMidEndSepPunct{\mcitedefaultmidpunct}
{\mcitedefaultendpunct}{\mcitedefaultseppunct}\relax
\EndOfBibitem
\bibitem[Qiao et~al.(2010)Qiao, Yang, Feng, Tse, Ding, Yao, Wang, and
  Niu]{Qiao:2010}
Qiao,~Z.; Yang,~S.~A.; Feng,~W.; Tse,~W.-K.; Ding,~J.; Yao,~Y.; Wang,~J.;
  Niu,~Q. Quantum Anomalous Hall Effect in Graphene from Rashba and Exchange
  Effects. \emph{Phys. Rev. B} \textbf{2010}, \emph{82}, 161414\relax
\mciteBstWouldAddEndPuncttrue
\mciteSetBstMidEndSepPunct{\mcitedefaultmidpunct}
{\mcitedefaultendpunct}{\mcitedefaultseppunct}\relax
\EndOfBibitem
\bibitem[Tse et~al.(2011)Tse, Qiao, Yao, MacDonald, and Niu]{Tse:2011}
Tse,~W.-K.; Qiao,~Z.; Yao,~Y.; MacDonald,~A.~H.; Niu,~Q. Quantum Anomalous Hall
  Effect in Single-Layer and Bilayer Graphene. \emph{Phys. Rev. B}
  \textbf{2011}, \emph{83}, 155447\relax
\mciteBstWouldAddEndPuncttrue
\mciteSetBstMidEndSepPunct{\mcitedefaultmidpunct}
{\mcitedefaultendpunct}{\mcitedefaultseppunct}\relax
\EndOfBibitem
\bibitem[Mauger and Godart(1986)Mauger, and Godart]{Mauger:1986}
Mauger,~A.; Godart,~C. The Magnetic, Optical, and Transport Properties of
  Representatives of a Class of Magnetic Semiconductors: The Europium
  Chalcogenides. \emph{Phys. Rep.} \textbf{1986}, \emph{141}, 51--176\relax
\mciteBstWouldAddEndPuncttrue
\mciteSetBstMidEndSepPunct{\mcitedefaultmidpunct}
{\mcitedefaultendpunct}{\mcitedefaultseppunct}\relax
\EndOfBibitem
\bibitem[Samsanov(1985)]{Samsanov:1982}
Samsanov,~G.~V., Ed. \emph{{T}he {O}xide {H}andbook}, 2nd ed.; {IFI}/{P}lenum:
  {N}ew {Y}ork, 1985\relax
\mciteBstWouldAddEndPuncttrue
\mciteSetBstMidEndSepPunct{\mcitedefaultmidpunct}
{\mcitedefaultendpunct}{\mcitedefaultseppunct}\relax
\EndOfBibitem
\bibitem[Oliver et~al.(1972)Oliver, Dimmock, McWhorter, and Reed]{Oliver:1972}
Oliver,~M.~R.; Dimmock,~J.~O.; McWhorter,~A.~L.; Reed,~T.~B. Conductivity
  Studies in Europium Oxide. \emph{Phys. Rev. B} \textbf{1972}, \emph{5},
  1078--1098\relax
\mciteBstWouldAddEndPuncttrue
\mciteSetBstMidEndSepPunct{\mcitedefaultmidpunct}
{\mcitedefaultendpunct}{\mcitedefaultseppunct}\relax
\EndOfBibitem
\bibitem[Schmehl et~al.(2007)Schmehl, Vaithyanathan, Herrnberger, Thiel,
  Richter, Liberati, Heeg, Rockerath, Kourkoutis, Muhlbauer, Boni, Muller,
  Barash, Schubert, Idzerda, Mannhart, and Schlom]{Schmehl:2007}
Schmehl,~A.; Vaithyanathan,~V.; Herrnberger,~A.; Thiel,~S.; Richter,~C.;
  Liberati,~M.; Heeg,~T.; Rockerath,~M.; Kourkoutis,~L.~F.; Muhlbauer,~S.
  et~al.  Epitaxial Integration of the Highly Spin-Polarized Ferromagnetic
  Semiconductor EuO with Silicon and GaN. \emph{Nat. Mater.}
  \textbf{2007}, \emph{6}, 882--887\relax
\mciteBstWouldAddEndPuncttrue
\mciteSetBstMidEndSepPunct{\mcitedefaultmidpunct}
{\mcitedefaultendpunct}{\mcitedefaultseppunct}\relax
\EndOfBibitem
\bibitem[Steeneken(2002)]{Steeneken:2002}
Steeneken,~P.~G. {N}ew Light on EuO Thin Films. Ph.D.\ thesis, {U}niversity of
  {G}roningen, 2002\relax
\mciteBstWouldAddEndPuncttrue
\mciteSetBstMidEndSepPunct{\mcitedefaultmidpunct}
{\mcitedefaultendpunct}{\mcitedefaultseppunct}\relax
\EndOfBibitem
\bibitem[Ulbricht et~al.(2008)Ulbricht, Schmehl, Heeg, Schubert, and
  Schlom]{Ulbricht:2008}
Ulbricht,~R.~W.; Schmehl,~A.; Heeg,~T.; Schubert,~J.; Schlom,~D.~G.
  Adsorption-Controlled Growth of EuO by Molecular-Beam Epitaxy. \emph{Appl.
  Phys. Lett.} \textbf{2008}, \emph{93}, 102105\relax
\mciteBstWouldAddEndPuncttrue
\mciteSetBstMidEndSepPunct{\mcitedefaultmidpunct}
{\mcitedefaultendpunct}{\mcitedefaultseppunct}\relax
\EndOfBibitem
\bibitem[Sutarto et~al.(2009)Sutarto, Altendorf, Coloru, Moretti~Sala,
  Haupricht, Chang, Hu, Sch\"u\ss{}ler-Langeheine, Hollmann, Kierspel, Hsieh,
  Lin, Chen, and Tjeng]{Sutarto:2009}
Sutarto,~R.; Altendorf,~S.~G.; Coloru,~B.; Moretti~Sala,~M.; Haupricht,~T.;
  Chang,~C.~F.; Hu,~Z.; Sch\"u\ss{}ler-Langeheine,~C.; Hollmann,~N.;
  Kierspel,~H. et~al.  Epitaxial and Layer-By-Layer Growth of EuO Thin Films on
  Yttria-Stabilized Cubic Zirconia (001) Using MBE Distillation. \emph{Phys.
  Rev. B} \textbf{2009}, \emph{79}, 205318\relax
\mciteBstWouldAddEndPuncttrue
\mciteSetBstMidEndSepPunct{\mcitedefaultmidpunct}
{\mcitedefaultendpunct}{\mcitedefaultseppunct}\relax
\EndOfBibitem
\bibitem[Swartz et~al.(2010)Swartz, Ciraldo, Wong, Li, Han, Lin, Mack, Shi,
  Awschalom, and Kawakami]{Swartz:2010}
Swartz,~A.~G.; Ciraldo,~J.; Wong,~J. J.~I.; Li,~Y.; Han,~W.; Lin,~T.; Mack,~S.;
  Shi,~J.; Awschalom,~D.~D.; Kawakami,~R.~K. Epitaxial EuO Thin Films on GaAs.
  \emph{Appl. Phys. Lett.} \textbf{2010}, \emph{97}, 112509\relax
\mciteBstWouldAddEndPuncttrue
\mciteSetBstMidEndSepPunct{\mcitedefaultmidpunct}
{\mcitedefaultendpunct}{\mcitedefaultseppunct}\relax
\EndOfBibitem
\bibitem[Swartz et~al.(2012)Swartz, Wong, Pinchuk, and Kawakami]{Swartz:2012}
Swartz,~A.~G.; Wong,~J. J.~I.; Pinchuk,~I.~V.; Kawakami,~R.~K. TiO$_2$ as an
  Electrostatic Template for Epitaxial Growth of EuO on MgO(001) by Reactive
  Molecular Beam Epitaxy. \emph{J. of Appl. Phys.} \textbf{2012},
  \emph{111}, 083912\relax
\mciteBstWouldAddEndPuncttrue
\mciteSetBstMidEndSepPunct{\mcitedefaultmidpunct}
{\mcitedefaultendpunct}{\mcitedefaultseppunct}\relax
\EndOfBibitem
\bibitem[Henrich(1996)]{Henrich:1996}
Henrich,~V.~E.; Cox,~P.~A.   \emph{The Surface Science of Metal Oxides}, 1st ed.; Cambridge University Press:
  Cambridge, 1996\relax
\mciteBstWouldAddEndPuncttrue
\mciteSetBstMidEndSepPunct{\mcitedefaultmidpunct}
{\mcitedefaultendpunct}{\mcitedefaultseppunct}\relax
\EndOfBibitem
\bibitem[Ahn and Shafer(1970)Ahn, and Shafer]{Ahn:1970}
Ahn,~K.~Y.; Shafer,~M.~W. Relationship Between Stoichiometry and Properties of
  EuO Films. \emph{J. of Appl. Phys.} \textbf{1970}, \emph{41},
  1260--1262\relax
\mciteBstWouldAddEndPuncttrue
\mciteSetBstMidEndSepPunct{\mcitedefaultmidpunct}
{\mcitedefaultendpunct}{\mcitedefaultseppunct}\relax
\EndOfBibitem
\bibitem[Novoselov et~al.(2004)Novoselov, Geim, Morozov, Jiang, Zhang, Dubonos,
  Grigorieva, and Firsov]{Novoselov:2004}
Novoselov,~K.~S.; Geim,~A.~K.; Morozov,~S.~V.; Jiang,~D.; Zhang,~Y.;
  Dubonos,~S.~V.; Grigorieva,~I.~V.; Firsov,~A.~A. Electric Field Effect in
  Atomically Thin Carbon Films. \emph{Science} \textbf{2004}, \emph{306},
  666--669\relax
\mciteBstWouldAddEndPuncttrue
\mciteSetBstMidEndSepPunct{\mcitedefaultmidpunct}
{\mcitedefaultendpunct}{\mcitedefaultseppunct}\relax
\EndOfBibitem
\bibitem[Hao et~al.(2004)Hao, Wang, Wang, Ni, Wang, Koo, Shen, and Thong]{Hao:2010}
Hao,~Y.; Wang,~Y.; Wang,~L.; Ni,~Z.; Wang,~Z.;
  Wang,~R.; Koo,~C.~K.; Shen,~Z.; Thong,~J.~T.~L. Probing Layer Number and Stacking Order of Few-Layer Graphene by Raman Spectroscopy \emph{Small} \textbf{2010}, \emph{6},
  195--200\relax
\mciteBstWouldAddEndPuncttrue
\mciteSetBstMidEndSepPunct{\mcitedefaultmidpunct}
{\mcitedefaultendpunct}{\mcitedefaultseppunct}\relax
\EndOfBibitem
\bibitem[Ferrari et~al.(2006)Ferrari, Meyer, Scardaci, Casiraghi, Lazzeri,
  Mauri, Piscanec, Jiang, Novoselov, Roth, and Geim]{Ferrari:2006}
Ferrari,~A.~C.; Meyer,~J.~C.; Scardaci,~V.; Casiraghi,~C.; Lazzeri,~M.;
  Mauri,~F.; Piscanec,~S.; Jiang,~D.; Novoselov,~K.~S.; Roth,~S. et~al.  Raman
  Spectrum of Graphene and Graphene Layers. \emph{Phys. Rev. Lett.}
  \textbf{2006}, \emph{97}, 187401\relax
\mciteBstWouldAddEndPuncttrue
\mciteSetBstMidEndSepPunct{\mcitedefaultmidpunct}
{\mcitedefaultendpunct}{\mcitedefaultseppunct}\relax
\EndOfBibitem
\bibitem[Gupta et~al.(2006)Gupta, Chen, Joshi, Tadigadapa, and
  Eklund]{Gupta:2006}
Gupta,~A.; Chen,~G.; Joshi,~P.; Tadigadapa,~S.; Eklund, Raman Scattering from
  High-Frequency Phonons in Supported n-Graphene Layer Films. \emph{Nano
  Lett.} \textbf{2006}, \emph{6}, 2667--2673\relax
\mciteBstWouldAddEndPuncttrue
\mciteSetBstMidEndSepPunct{\mcitedefaultmidpunct}
{\mcitedefaultendpunct}{\mcitedefaultseppunct}\relax
\EndOfBibitem
\bibitem[Malard et~al.(2009)Malard, Pimenta, Dresselhaus, and
  Dresselhaus]{Malard:2009}
Malard,~L.; Pimenta,~M.; Dresselhaus,~G.; Dresselhaus,~M. Raman Spectroscopy in
  Graphene. \emph{Phys. Rep.} \textbf{2009}, \emph{473}, 51--87\relax
\mciteBstWouldAddEndPuncttrue
\mciteSetBstMidEndSepPunct{\mcitedefaultmidpunct}
{\mcitedefaultendpunct}{\mcitedefaultseppunct}\relax
\EndOfBibitem
\bibitem[Tang et~al.(2010)Tang, Guoxin, and Gao]{Tang:2010}
Tang,~B.; Guoxin,~H.; Gao,~H. Raman Spectroscopic Characterization of Graphene.
  \emph{Appl. Spectrosc. Rev.} \textbf{2010}, \emph{45}, 369--407\relax
\mciteBstWouldAddEndPuncttrue
\mciteSetBstMidEndSepPunct{\mcitedefaultmidpunct}
{\mcitedefaultendpunct}{\mcitedefaultseppunct}\relax
\EndOfBibitem
\bibitem[Pisana et~al.(2007)Pisana, Lazzeri, Casiraghi, Geim, Ferrari, and
  Mauri]{Pisana:2007}
Pisana,~S.; Lazzeri,~M.; Casiraghi,~K.~S.,~C.~Novoselov; Geim,~A.~K.;
  Ferrari,~A.~C.; Mauri,~F. Breakdown of the Adiabatic Born-Oppenheimer
  Approximation in Graphene. \emph{Nat. Mater.} \textbf{2007}, \emph{6},
  198\relax
\mciteBstWouldAddEndPuncttrue
\mciteSetBstMidEndSepPunct{\mcitedefaultmidpunct}
{\mcitedefaultendpunct}{\mcitedefaultseppunct}\relax
\EndOfBibitem
\bibitem[Das et~al.(2008)Das, Pisana, Chakraborty, Piscanec, Saha, Waghmare,
  Novoselov, Krishnamurthy, Geim, Sood, and Ferrari]{Das:2008}
Das,~A.; Pisana,~S.; Chakraborty,~B.; Piscanec,~S.; Saha,~S.~K.;
  Waghmare,~U.~V.; Novoselov,~K.~S.; Krishnamurthy,~H.~R.; Geim,~A.~K.;
  Sood,~A.~K. et~al.  Monitoring Dopants by Raman Scattering in an
  Electrochemically Top-Gated Graphene Transistor. \emph{Nat. Nanotechnol.}
  \textbf{2008}, \emph{3}, 210\relax
\mciteBstWouldAddEndPuncttrue
\mciteSetBstMidEndSepPunct{\mcitedefaultmidpunct}
{\mcitedefaultendpunct}{\mcitedefaultseppunct}\relax
\EndOfBibitem
\bibitem[Ni et~al.(2008)Ni, Wang, Ma, Kasim, Wu, and Shen]{Ni:2008a}
Ni,~Z.~H.; Wang,~H.~M.; Ma,~Y.; Kasim,~J.; Wu,~Y.~H.; Shen,~Z.~X. Tunable
  Stress and Controlled Thickness Modification in Graphene by Annealing.
  \emph{ACS Nano} \textbf{2008}, \emph{2}, 1033--1039\relax
\mciteBstWouldAddEndPuncttrue
\mciteSetBstMidEndSepPunct{\mcitedefaultmidpunct}
{\mcitedefaultendpunct}{\mcitedefaultseppunct}\relax
\EndOfBibitem
\bibitem[Ni et~al.(2008)Ni, Yu, Lu, Wang, Feng, and Shen]{Ni:2008b}
Ni,~Z.~H.; Yu,~T.; Lu,~Y.~H.; Wang,~Y.~Y.; Feng,~Y.~P.; Shen,~Z.~X. Uniaxial
  Strain on Graphene: Raman Spectroscopy Study and Band-Gap Opening. \emph{ACS
  Nano} \textbf{2008}, \emph{2}, 2301--2305\relax
\mciteBstWouldAddEndPuncttrue
\mciteSetBstMidEndSepPunct{\mcitedefaultmidpunct}
{\mcitedefaultendpunct}{\mcitedefaultseppunct}\relax
\EndOfBibitem
\bibitem[Ding et~al.(2010)Ding, Ji, Chen, Herklotz, D\"orr, Mei, Rastelli, and
  Schmidt]{Ding:2010}
Ding,~F.; Ji,~H.; Chen,~Y.; Herklotz,~A.; D\"orr,~K.; Mei,~Y.; Rastelli,~A.;
  Schmidt,~O.~G. Stretchable Graphene: A Close Look at Fundamental Parameters
  through Biaxial Straining. \emph{Nano Lett.} \textbf{2010}, \emph{10},
  3453--3458\relax
\mciteBstWouldAddEndPuncttrue
\mciteSetBstMidEndSepPunct{\mcitedefaultmidpunct}
{\mcitedefaultendpunct}{\mcitedefaultseppunct}\relax
\EndOfBibitem
\bibitem[Cheng et~al.(2011)Cheng, Zhou, Wang, Li, Wang, and Fang]{Cheng:2011}
Cheng,~Z.; Zhou,~Q.; Wang,~C.; Li,~Q.; Wang,~C.; Fang,~Y. Toward Intrinsic
  Graphene Surfaces: A Systematic Study on Thermal Annealing and Wet-Chemical
  Treatment of SiO$_2$-Supported Graphene Devices. \emph{Nano Lett.}
  \textbf{2011}, \emph{11}, 767--771\relax
\mciteBstWouldAddEndPuncttrue
\mciteSetBstMidEndSepPunct{\mcitedefaultmidpunct}
{\mcitedefaultendpunct}{\mcitedefaultseppunct}\relax
\EndOfBibitem
\bibitem[Pi et~al.(2009)K. Pi and K. M. McCreary and W. Bao and Wei Han and Y. F. Chiang and Yan Li and S. W. Tsai and C. N. Lau and R. K. Kawakami]{Pi:2009}
Pi,~K.; McCreary,~K. M.; Bao,~W.; Han,~W.; Chiang,~Y. F.; Li,~Y.; Tsai,~S. W.; Lau,~C. N.; Kawakami,~R. K.  Electronic Doping and Scattering by Transition Metals on Graphene. \emph{Phys. Rev. B}
  \textbf{2009}, \emph{80}, 075406\relax
\mciteBstWouldAddEndPuncttrue
\mciteSetBstMidEndSepPunct{\mcitedefaultmidpunct}
{\mcitedefaultendpunct}{\mcitedefaultseppunct}\relax
\EndOfBibitem
\bibitem[Li et~al.(2009)Li, Cai, An, Kim, Nah, Yang, Piner, Velamakanni, Jung,
  Tutuc, Banerjee, Colombo, and Ruoff]{Li:2009}
Li,~X.; Cai,~W.; An,~J.; Kim,~S.; Nah,~J.; Yang,~D.; Piner,~R.;
  Velamakanni,~A.; Jung,~I.; Tutuc,~E. et~al.  Large-Area Synthesis of
  High-Quality and Uniform Graphene Films on Copper Foils. \emph{Science}
  \textbf{2009}, \emph{324}, 1312--1314\relax\mciteBstWouldAddEndPuncttrue
\mciteSetBstMidEndSepPunct{\mcitedefaultmidpunct}
{\mcitedefaultendpunct}{\mcitedefaultseppunct}\relax
\EndOfBibitem


\end{mcitethebibliography}
\end{document}